\documentclass{article}

\PassOptionsToPackage{numbers, compress}{natbib}

\usepackage[preprint]{neurips_2024}

\usepackage[utf8]{inputenc} 
\usepackage[T1]{fontenc}    
\usepackage{hyperref}       
\usepackage{url}            
\usepackage{booktabs}       
\usepackage{amsfonts}       
\usepackage{nicefrac}       
\usepackage{microtype}      
\usepackage{xcolor}         
\usepackage{graphicx}

\title{Optimising EEG decoding with refined sampling and multimodal feature integration}

\author{%
  Arash Akbarinia\thanks{The source code all experimental material available at \url{https://github.com/ArashAkbarinia/ides} .} \\
  Department of Experimental Psychology\\
  University of Giessen, Germany\\
  \\
  \texttt{arash.akbarinia@psychol.uni-giessen.de} \\
}

\begin{document}

\maketitle

\begin{abstract}
Electroencephalography (EEG) is a neuroimaging technique that records brain neural activity with high temporal resolution. Unlike other methods, EEG does not require prohibitively expensive equipment and can be easily set up using commercially available portable EEG caps, making it an ideal candidate for brain-computer interfaces. However, EEG signals are characterised by poor spatial resolution and high noise levels, complicating their decoding. In this study, we employ a contrastive learning framework to align encoded EEG features with pretrained CLIP features, achieving a 7\% improvement over the state-of-the-art in EEG decoding of object categories. This enhancement is equally attributed to (1) a novel online sampling method that boosts the signal-to-noise ratio and (2) multimodal representations leveraging visual and language features to enhance the alignment space. Our analysis reveals a systematic interaction between the architecture and dataset of pretrained features and their alignment efficacy for EEG signal decoding. This interaction correlates with the generalisation power of the pretrained features on ImageNet-O/A datasets ($r=.5$). These findings extend beyond EEG signal alignment, offering potential for broader applications in neuroimaging decoding and generic feature alignments.
\end{abstract}

\section{Introduction}

The quest to unravel the functions of the human brain has long captivated neuroscientists \cite{miyawaki2008visual}. Traditionally, invasive methods, such as implanting electrodes in animal brains, have been employed to record neuronal activities \cite{cogan2008neural}. These techniques have yielded significant insights into mammalian brain functions \cite{hubel1962receptive}, but their applicability to humans is severely restricted. To circumvent these limitations, several noninvasive neuroimaging techniques have been developed, allowing the recording of neuronal activity in the human brain. Among these techniques, functional magnetic resonance imaging (fMRI) has garnered considerable attention due to its high spatial resolution and relatively low noise levels \cite{takagi2023high, scotti2024reconstructing}. Unlike positron emission tomography (PET), fMRI does not involve radioactivity exposure. However, the high costs, bulky equipment, and low temporal resolution of fMRI limit its practical applications.

Magnetoencephalography (MEG) offers an alternative with high temporal resolution, yet it shares the drawbacks of expensive and cumbersome equipment, restricting its usability in real-world settings. In contrast, electroencephalography (EEG) presents a more accessible option \cite{contini2017decoding}. EEG is cost-effective, with caps available for a few hundred Euros, and its lightweight, portable nature makes it suitable for various applications. Despite its high temporal resolution, EEG is hampered by significant noise and poor spatial resolution. EEG recordings are notably sensitive to artefacts caused by minor movements, such as blinking and yawning. This raises critical questions about the extent of useful information contained within EEG signals and their potential for practical applications, particularly in brain-computer interfaces (BCIs) \cite{kamitani2005decoding, kay2008identifying}.

BCIs represent a promising frontier for EEG application, enabling direct communication between the brain and external devices \cite{gao2021interface}. This technology holds potential for numerous applications, including aiding individuals with disabilities, enhancing cognitive functions, and providing new methods of human-computer interaction. Despite these promising prospects, the efficacy of current EEG-based BCI systems is limited by the quality of the EEG signals and the methods used to decode them.

A standard benchmark for evaluating the efficacy of neuroimaging techniques, including EEG, is object recognition. This involves determining the ability to identify and categorise visual stimuli based on recorded brain activity. Recently, the THINGS EEG2 Dataset \cite{gifford2022large} was introduced, recording EEG signals from ten participants as they viewed natural images representing thousands of different concepts. This extensive dataset provides a valuable resource for examining the concept-specific information retained in EEG signals and facilitates the use of data-intensive machine learning techniques. Several studies leveraging this dataset have demonstrated that object recognition accuracy using deep learning techniques significantly surpasses chance levels, highlighting the potential of EEG in complex cognitive tasks.

Research efforts have employed various innovative approaches to improve EEG decoding performance. For instance, Du et al. \cite{duetal} incorporated multimodal brain-visual-language features, enhancing the interpretability and robustness of the decoded signals. Similarly, Song et al. \cite{nice}  proposed a contrastive learning alignment framework to improve EEG decoding performance by aligning it with pretrained features, thus leveraging the strengths of different modalities. Additionally, Li et al. \cite{atm} demonstrated that these aligned features could be used to generate images directly from EEG signals, showcasing the potential of advanced decoding techniques.

Despite these advancements, the current performance of state-of-the-art methods remains inadequate for practical real-world applications of BCIs. A significant challenge faced by previous works is the overfitting of neural networks to training data, resulting in poor generalisation to new, unseen data \cite{schirrmeister2017deep}. Addressing this issue is crucial for developing reliable and effective EEG-based decoding systems.

In this paper, we propose two novel techniques to mitigate the overfitting problem in EEG decoding of object recognition. First, we introduce a new sampling technique that expands the training space by sampling EEG signals across both repetitions and concept dimensions. This approach aims to diversify the training data, thereby improving the network's ability to generalise. Second, we augment the alignment space by incorporating multimodal features from both the visual and language domains. By enriching the feature space, we aim to enhance the robustness and accuracy of the decoded signals. While the proposed framework does not completely eliminate overfitting, it significantly reduces this problem, leading to significant improvements in test accuracy.

\section{Method}

\subsection{Refined sampling}

EEG data is notoriously noisy, necessitating strategies to increase the signal-to-noise ratio (SNR) \cite{pan2022matt}. Traditionally, this is achieved by averaging EEG recordings across several repetitions of the same stimulus. While effective, this approach is limited by the small number of repetitions for an identical stimulus. To address this, we propose expanding the averaging process to include similar stimuli, defined as identical concepts, during the training stage. This not only enhances the SNR but also increases the EEG sampling space. We refer to our method as \textbf{InterDimensional EEG Sampling (IDES)}.

We applied IDES to the THINGS EEG2 Dataset \cite{gifford2022large}, which includes four repetitions per image and ten images per concept, resulting in a total sampling space of 40 elements. At each iteration, we randomly select seven of these elements and average the EEG signals across them. The rationale behind this approach is that for object recognition decoding, it is irrelevant whether the participant observed a specific type of tree or snail; what matters is that they viewed an instance of a tree or snail.

IDES differs from standard EEG augmentation techniques that artificially modify the signal within the same trial, either by removing parts of the signal or interpolating data. The advantage of IDES is that it extends beyond a single trial, sampling from the true distribution of real-world stimuli. This method not only improves the SNR but also provides a more robust and generalized representation of EEG signals associated with object recognition.

\subsection{Multimodal EEG decoding framework}

\begin{figure}
    \centering
    \includegraphics[width=\textwidth]{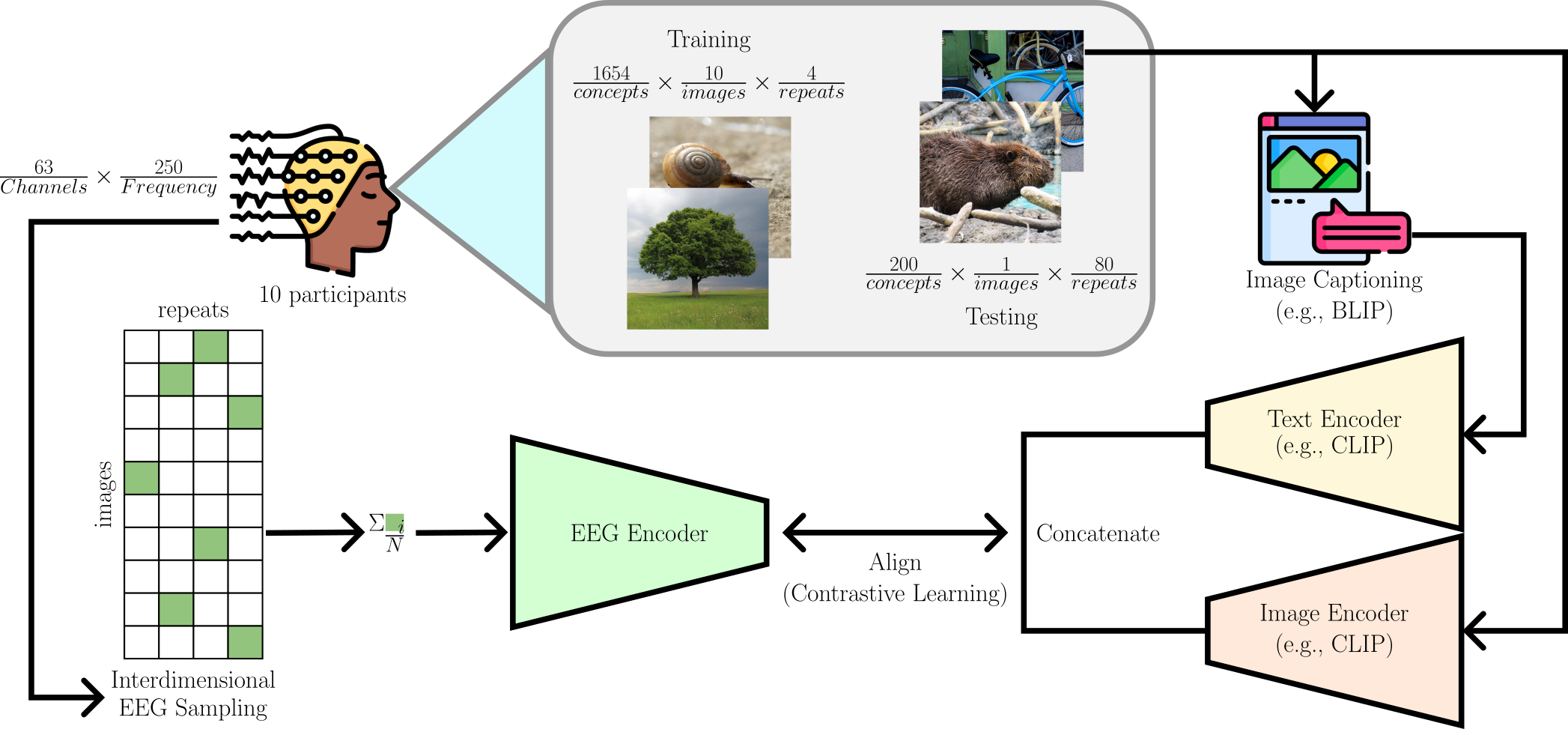}
    \caption{The schematic flowchart illustrates the proposed framework for decoding EEG signals by aligning them with multimodal pretrained features. During training, the EEG signals are sampled across images and repeat dimensions to expand the EEG training space. At test time, EEG signals are averaged across all repeats. An EEG encoder extracts EEG features, which are then aligned with features extracted from pretrained networks using contrastive learning. These pretrained features are obtained by concatenating text and image feature vectors. Text features are generated by feeding images into an image captioning network to produce descriptive texts, which are subsequently processed by a text encoder. Image features are directly extracted by an image encoder.}
    \label{fig:flowchart}
\end{figure}

To decode EEG signals for object recognition, we modified the alignment framework proposed in NICE. The schematic of our decoding framework is depicted in Figure \ref{fig:flowchart}. EEG signals are first processed using the InterDimensional EEG Sampling (IDES) method, as described above. These processed signals are then input into an EEG Encoder, which outputs an EEG feature vector of size $X$, depending on the size of the pretrained features.

We extract two types of pretrained features from the images observed by the participants. Similar to NICE, we use image encoders to extract visual features from these images. Additionally, we employ an image captioning network (i.e., BLIP \cite{li2022blip}) to generate descriptive text, limited to a maximum of 20 words per image. The rationale behind this approach is that the recorded EEG signal reflects not only the activation of neurons in the visual areas of the brain but also in other regions, particularly the language areas \cite{chan2011decoding}. We hypothesise that incorporating language features is crucial, as these areas are likely involved in processing the observed images.

The generated captions are subsequently passed through a text encoder to extract text features. These text features are then concatenated with the visual features to form a comprehensive multimodal feature representation. The EEG features are compared to these pretrained features using a contrastive learning loss function, which computes the pairwise cosine similarity distance. This approach aims to align the EEG signals with the multimodal pretrained features, leveraging both visual and language information to improve the predictive power of object recognition from EEG data.

\subsection{Pretrained features}

We adopted an offline feature extraction approach to enhance training efficiency, extracting features from both the text and image encoders. Prior to inputting them into pretrained networks, images were resized to dimensions of $224 \times 244$. Visual features were acquired directly by forwarding the images through an image encoder. Conversely, language features were obtained through a two-step process: first, images were fed into an image captioning network (BLIP \cite{li2022blip}), generating textual descriptions with a maximum length of 20 words; subsequently, these captions were processed by a language encoder to extract language features.

For the test concepts, we generated two distinct sets of features: 
\begin{enumerate}
    \item \textbf{In-Distribution (ID)} is obtained directly by passing the test images to pretrained networks.
    \item \textbf{Out-of-Distribution (OOD)} is obtained by excluding the test images observed by EEG participants from the THINGS Dataset \cite{hebart2019things}. For each concept, all remaining images (about 13 images per concept) were forwarded through the pretrained networks, and the resulting features were averaged to create the final pretrained features.
\end{enumerate}

\subsection{EEG Encoder}

We examined several EEG decoder architectures, including NICE \cite{nice}, ATM \cite{atm}, EEGNet \cite{lawhern2018eegnet}, EEGInception \cite{santamaria2020eeg}, and EEGResNet \cite{HBM:HBM23730}. While NICE and ATM significantly outperformed the other architectures, we did not observe any substantial difference between the performance of NICE and ATM. Consequently, we have chosen to use the ATM architecture for the results reported in the main manuscript. However, results obtained using the NICE architecture are available in the supplementary materials, where a comprehensive spreadsheet of all experimental results is provided.

\section{Experiments}

\subsection{THINGS EEG2 Dataset}

For our experimental investigations, we utilised the THINGS EEG2 Dataset \cite{gifford2022large}, encompassing EEG signals obtained from ten participants while viewing natural images sourced from the THINGS Dataset. Notably, concepts within the THINGS EEG2 dataset are disjointed, ensuring that identical concepts do not feature in both the training and test sets. The training set comprises 1654 concepts, each associated with 10 images and subjected to 4 repetitions, whereas the test set consists of 200 concepts, each with a single image subjected to 80 repetitions. The raw EEG signal is composed of 63 channels and sampled at a frequency of 1000 Hz. Prior to analysis, we preprocessed the raw data, downsampling it to 250 Hz and segmenting it into trials spanning from 0 to 1000 ms following stimulus onset. Multivariate noise normalisation was exclusively applied to the training data. Additionally, baseline correction was executed using the mean of the 200 ms pre-stimulus data. The input dimensions of our EEG Encoder consist of a matrix measuring $63 \times 250$.

\subsection{Training and testing}

The proposed framework is implemented in PyTorch, ensuring consistency across training and testing settings to mitigate the influence of random variables. To establish a validation set, the last image of each concept is excluded from training. All EEG Encoders undergo training for 40 epochs, encompassing 14,886 images. We employ the AdamW optimiser with a learning rate of $3e-4$ and $\beta =(0.9, 0.999)$. No data augmentation is applied to the input EEG signals. Training involves two regimes: 
\begin{enumerate}
    \item \textbf{Intraparticipant}: Training and testing on data from the same participant.
    \item \textbf{Interparticipant}: Training on data from nine participants and testing on the left-out 
\end{enumerate}
We evaluate the networks using both In-Distribution (ID) and Out-of-Distribution (OOD) test sets. While the ID results, which exhibit notably higher accuracy (7-10\%), are relegated to the appendices, we present the OOD results in the main manuscript, as they signify the true generalisation capabilities of the trained networks \cite{geirhos2020shortcut, akbarinia2019paradox}. We report both Top-1 and Top-5 accuracies, with the chance level set at 1 out of 200.

\subsection{Baseline comparison}

\begin{figure}
    \centering
    \includegraphics[width=\textwidth]{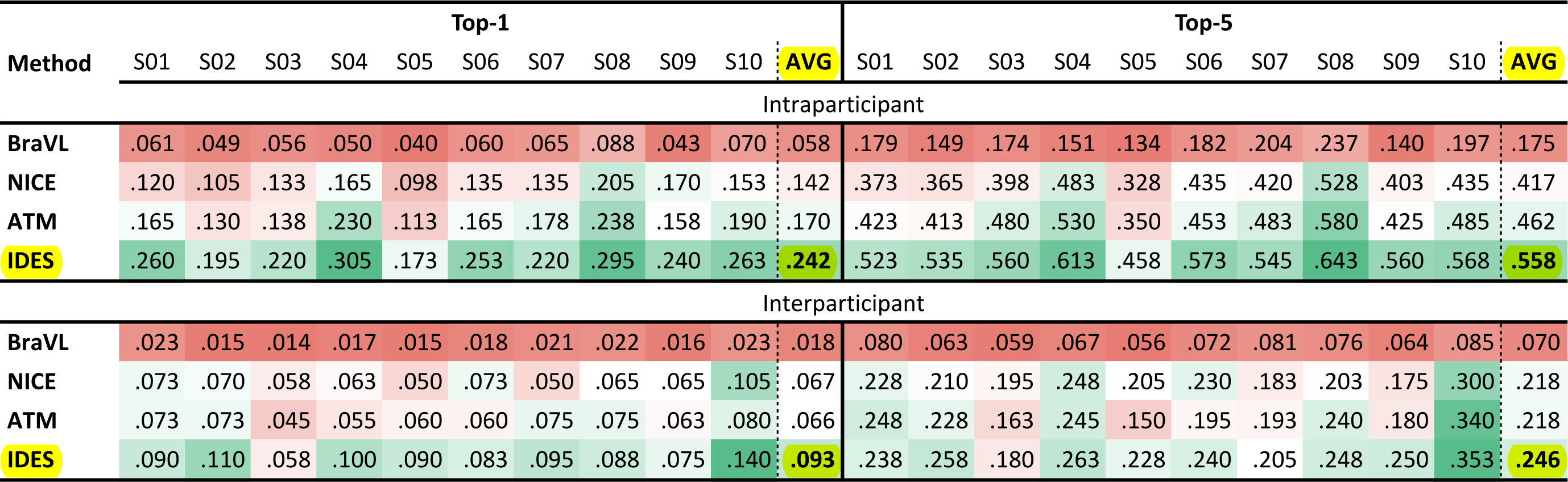}
    \caption{This table presents the out-of-distribution (OOD) classification accuracy across 200 test concepts from the THINGS EEG2 Dataset \cite{gifford2022large}. Each cell is colour-coded from worst (red) to best (green) to visualise performance. The proposed model is highlighted in yellow. Results are shown for individual participants (S01 to S10), and AVG represents the average accuracy across all ten participants. Intraparticipant refers to training and testing an EEG encoder on data from the same participant. Interparticipant refers to training on data from nine participants and testing on the left-out participant.}
    \label{fig:baseline}
\end{figure}

We commence by comparing the proposed framework (IDES) against three state-of-the-art methods (BraVL \cite{duetal}, NICE \cite{nice}, and ATM \cite{atm}), with their corresponding classification accuracies delineated in Figure \ref{fig:baseline}. The proposed framework demonstrates a substantial enhancement over the state-of-the-art, elevating Top-1 accuracy by approximately 7\% (intraparticipant) and 3\% (interparticipant). Notably, this improvement persists at the participant level. Taken together, these findings underscore the efficacy of interdimensional EEG sampling and multimodal feature alignment in boosting EEG decoding performance. Subsequently, we proceed to quantify the individual contribution of each component to the overall performance gain.

\begin{figure}
    \centering
    \includegraphics[width=\textwidth]{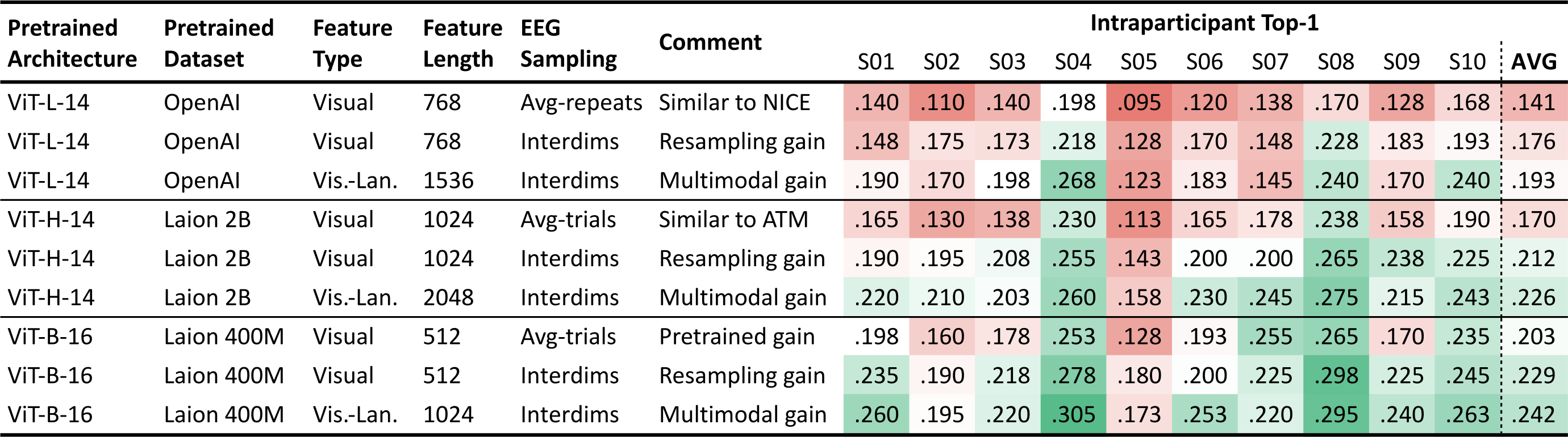}
    \caption{This table breaks down the performance gain of the proposed framework by comparing the intraparticipant classification accuracy across several settings. Each cell is colour-coded from worst (red) to best (green) to visualise performance. The baselines used are NICE and ATM, which sample EEG signals by averaging across repeats and aligning EEG features to pretrained visual features. Each baseline is improved through two methods: (1) expanding the EEG space using interdimensional sampling, and (2) aligning EEG features to multimodal visual-language features.}
    \label{fig:gains}
\end{figure}

To dissect the contribution of each component to the observed performance enhancement, we trained networks that solely varied in one aspect and compared the resulting accuracies (Figure \ref{fig:gains}). The 7\% performance uplift achieved in our framework emanates from three distinct factors. 

\begin{itemize}
    \item The most substantial performance gain arises from the proposed interdimensional EEG sampling, accounting for more than 3.5\% improvement regardless of the pretrained architecture or the baseline method (NICE or ATM). 
    \item integrating multimodal features (visual-language) into the alignment process yields a 1.5\% performance boost, again independent of the pretrained architecture or the baseline method (NICE or ATM).
    \item The residual performance gain is attributed to certain pretrained features exhibiting superior generalisation to EEG decoding. We will delve into this further, noting that the dominant factor influencing this gain is the pretrained dataset rather than the pretrained architecture.
\end{itemize}

In summary, interparticipant Top-1 accuracy experiences a 5\% increase with the adoption of the proposed interdimensional EEG sampling method and multimodal feature alignment. This improvement holds irrespective of employing the NICE and ATM baselines or leveraging the best-discovered pretrained features (e.g., OpenCLIP ViT-B-16 trained on Laion 400M \cite{cherti2023reproducible}). Next, we will explore the effect of each factor through statistical analysis by using the Wilcoxon Signed-Rank Test.

\subsection{EEG sampling effect}

\begin{figure}
    \centering
    \includegraphics[width=\textwidth]{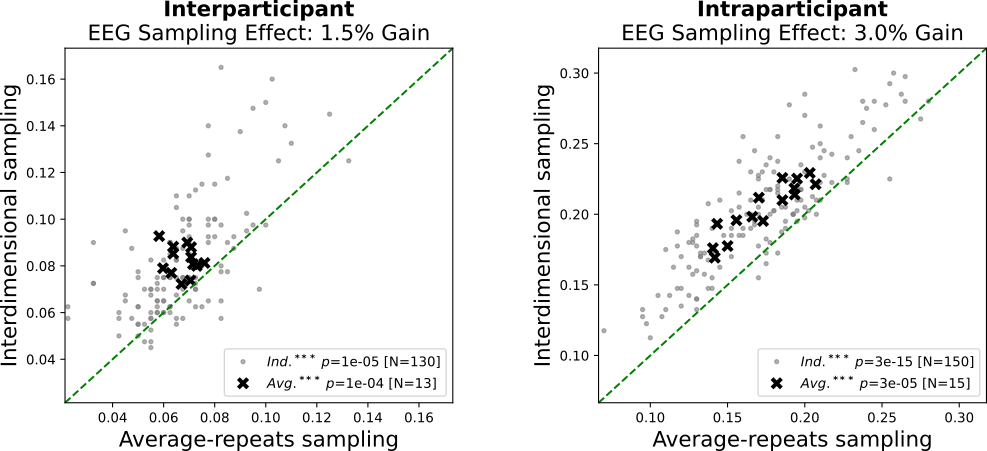}
    \caption{This figure demonstrates the effect of EEG sampling by comparing the classification accuracy of pairs of networks that are identical in all aspects (e.g., architecture, pretrained networks and features) except for their EEG sampling method. Each data point represents the Top-1 accuracy for individual participants, shown as grey-filled circles. Bold black crosses indicate the average accuracy across participants for a given test setting. The dashed green line represents the identity line; points above this diagonal line indicate a performance gain by using interdimensional sampling. The number of data points, p-values, and their significance are indicated in the legend.}
    \label{fig:sampling_effect}
\end{figure}

To systematically assess the impact of the proposed EEG sampling method, we conducted training sessions for 26 interparticipant and 30 intraparticipant networks, employing various pretrained features and architectures. Half of these networks were trained using the conventional average-repeats sampling method, while the remaining half utilised the proposed interdimensional method, with all other parameters held constant. Figure \ref{fig:sampling_effect} presents a comparative analysis of these two sets of networks, plotting the Top-1 accuracy of interdimensional sampling against that of average-repeats sampling. Data points aligning with the identity line (indicated by a dashed green line) suggest no discernible difference between the two sampling methods, whereas points situated in the upper triangle denote a performance boost by interdimensional sampling, and those in the lower triangle indicate a boost by average-repeat sampling.

Our observations reveal that the proposed sampling method enhances interparticipant accuracy by 1.5\% and intraparticipant accuracy by 3\%. To ascertain the statistical significance of these performance gains, we conducted Wilcoxon Signed-Rank Tests \cite{woolson2007wilcoxon}, both at the aggregate level and, notably, at the individual participant level. This dual-level analysis is crucial as it ensures that observed improvements in average results are consistent at the individual participant level. Our analysis yielded $p$-values significantly smaller than $\leq 0.001$ for both interparticipant and intraparticipant conditions in both individual and aggregate analyses. These findings provide compelling evidence that the proposed interdimensional sampling method significantly enhances EEG decoding performance.

\subsection{Multimodal effect}

\begin{figure}
    \centering
    \includegraphics[width=\textwidth]{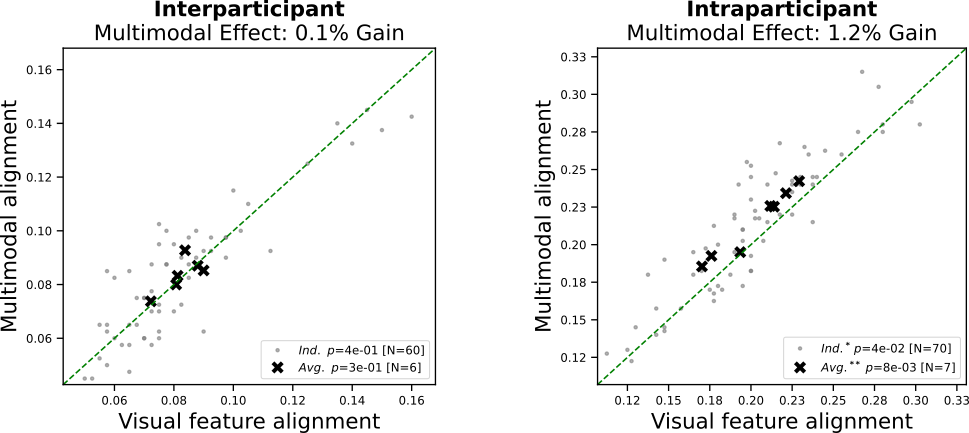}
    \caption{This figure demonstrates the effect of multimodal feature alignment by comparing the classification accuracy of pairs of networks that are identical in all aspects (e.g., architecture, sampling, and pretrained network) except for their pretrained features. Each data point represents the Top-1 accuracy for individual participants, shown as grey-filled circles. Bold black crosses indicate the average accuracy across participants for a given test setting. The dashed green line represents the identity line; points above this diagonal line indicate a performance gain by using multimodal (visual-language) pretrained features. The number of data points, p-values, and their significance are indicated in the legend.}
    \label{fig:multimodal_effect}
\end{figure}

To systematically evaluate the impact of the proposed multimodal feature alignment, we conducted training sessions for 12 interparticipant and 14 intraparticipant networks, employing various pretrained features and architectures. Half of these networks were trained solely using visual features, while the remaining half incorporated the proposed multimodal visual-language features. Similar to the previous analysis, we visualised these results using paired comparison analysis and conducted statistical analysis at both the individual participant and average levels (Figure \ref{fig:multimodal_effect}). Our findings reveal that the incorporation of multimodal features yields a marginal performance gain (0.1\%) for interparticipant networks, which, however, is not statistically significant. Conversely, intraparticipant networks exhibit a notable 1.2\% performance enhancement through visual-language feature alignment, which proves statistically significant at both individual and average levels, albeit with moderate significance at the individual level.

The discrepancy in the efficacy of incorporating language features between interparticipant and intraparticipant networks may stem from greater inter-individual variability in language-related brain areas \cite{tanner2014erps}, compared to the more uniform nature of visual areas. In our current framework, visual and language features are concatenated into a single vector, and EEG features are aligned with this composite representation, disregarding channel-specific information. A more nuanced approach for future research could involve aligning language-related EEG channels with pretrained language features and visual-related EEG channels with pretrained visual features. This targeted alignment may better capture the distinct neural representations of visual and language information.

Additionally, future investigations could explore the use of different pretrained networks for extracting visual and language features. In the current study, we used the same pretrained network to obtain both visual and language features. However, examining whether leveraging multimodal information from different pretrained networks could better democratise the pretrained feature space and further enhance decoding performance would be valuable. These proposed avenues offer promising directions for improving the integration of multimodal features in EEG-based decoding systems.

\subsection{Pretrained effect}

\begin{figure}
    \centering
    \includegraphics[width=\textwidth]{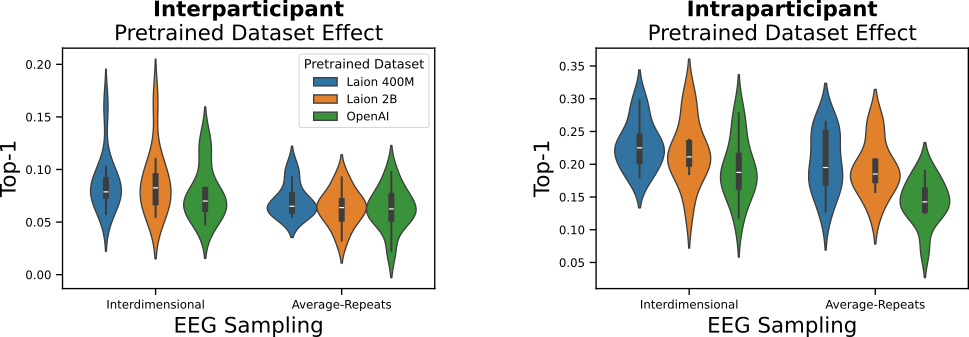}
    \caption{This figure demonstrates the effect of a pretrained dataset for an identical pretrained architecture (ViT-B-16) using only the visual features. The violin plot shows the distribution of Top-1 accuracy across ten participants.}
    \label{fig:pretrained_effect}
\end{figure}

It was previously reported that CLIP features \cite{radford2021learning} outperform ImageNet features \cite{deng2009imagenet} in EEG decoding alignment \cite{nice}. Our findings corroborate this observation. We further analysed several CLIP networks (from both OpenAI and OpenCLIP), noting that the type of pretrained dataset used to train the CLIP network significantly affects performance. Figure \ref{fig:pretrained_effect} compares three different pretrained datasets (Laion 400M, Laion 2B, and OpenAI) used to train an identical architecture (ViT-B-16). Overall, Laion 400M \cite{schuhmann2021laion} consistently results in the highest EEG decoding power, regardless of the sampling method (interdimensional or average-repeats) or the training regime (interparticipant or intraparticipant).

Although we could not perform a statistical analysis due to the limited number of pretrained networks available for all three datasets, the trend remains clear across our experiments. We can definitively rule out dataset size as the cause of performance differences, as Laion 2B performs worse than Laion 400M despite being larger. This discrepancy is also unrelated to the average zero-shot accuracy of these pretrained networks on 38 different datasets reported by OpenCLIP. However, we observed a high correlation ($r=0.5$) with the performance on ImageNet-O/A datasets \cite{hendrycks2021nae}, which contain images not included in the original ImageNet and are used to test the generalisation power of pretrained networks. Thus, it appears that pretrained features that generalise well to these datasets also result in better alignment in the EEG decoding problem.

\section{Discussion}

In this manuscript, we proposed two key innovations: (1) a novel EEG sampling method that collapses EEG data across the concept and repetitions dimensions, and (2) the incorporation of language features into a multimodal feature alignment process. These combined approaches resulted in a 5\% performance improvement across examined baselines, irrespective of other factors. Statistical analysis confirmed that the performance gains from the novel sampling method are statistically significant, as are the gains from multimodal features in the case of intraparticipant networks.  Additionally, we achieved a 2\% improvement over the state-of-the-art by exploring different pretrained features, finding that the choice of pretrained dataset was the underlying cause of this improvement. We believe the findings of this manuscript are sufficiently general to be directly applicable to other neuroimaging decoding tasks, including MEG and fMRI. Lastly, our results demonstrate that it is possible to achieve nearly 25\% accuracy at the intraparticipant level, indicating that there is substantial object-specific information in EEG signals that can be harnessed for various brain-computer interface applications.

\subsection{Computational cost}

The proposed interdimensional sampling method does not significantly increase the computational cost of the training procedure. Training on data from a single participant using the average-trial sampling method (our baseline) takes approximately five minutes on a GeForce 4090 GPU. The introduction of interdimensional sampling extends the computational time by just one minute, a cost that can be mitigated through optimisation of the random trial selection process.

Similarly, the use of multimodal visual-language pretrained features does not necessarily increase the computational cost, as this largely depends on the feature size. For instance, our best results are obtained with ViT-B-16 pretrained features, which have a length of 1024 (comprising 512 visual and 512 language features), matching the size of the visual pretrained features of the ViT-H-14 used in ATM. However, for the same architecture (ViT-B-16), using the visual-language features increases the training time from six to eight minutes.

\subsection{Limitations}

The most significant limitation of this manuscript is that the proposed framework has only been tested on a single EEG dataset. Consequently, it is crucial to determine whether the findings generalise to other EEG datasets, both for object recognition decoding and for other tasks. Additionally, as discussed in the multimodal effect section, although incorporating language features yields the highest decoding accuracy in both inter- and intraparticipant networks, the effect is not statistically significant in interparticipant networks.

\subsection{Broader impact}

While the findings of this paper may not have immediate societal impacts, advancements in EEG decoding hold significant potential for society, encompassing both positive and negative aspects \cite{niso2022good}. Improved EEG decoding can enhance medical applications by aiding in the diagnosis and treatment of neurological disorders, improving brain-computer interfaces (BCIs) for individuals with severe motor disabilities, facilitating cognitive enhancement tools, and aiding in the early detection and treatment of mental health conditions. However, these advancements also raise concerns, such as privacy issues related to the misuse of brain data, potential inequalities due to the high cost and limited accessibility of advanced EEG technologies, the risk of intrusive surveillance, and the psychological impact of being monitored. Therefore, while the benefits are substantial, ethical, privacy, and societal challenges must be carefully managed.

\section*{Acknowledgements}

This research was funded by the Deutsche Forschungsgemeinschaft SFB/TRR 135 (grant number 222641018) TP S.


\bibliographystyle{plain}
\bibliography{references.bib}






\end{document}